\documentclass[conference]{IEEEtran}
\usepackage[numbers,sort&compress]{natbib}

\usepackage[normalem]{ulem}
\usepackage{multirow}
\usepackage{multicol}
\usepackage{makecell}
\usepackage{boldline}
\usepackage[normalem]{ulem}
\usepackage{graphicx}
\usepackage{caption}
\usepackage{subfig}
\usepackage{xcolor}
\usepackage{listings}
\usepackage{xparse}

\NewDocumentCommand{\code}{v}{%
\texttt{\textcolor{black}{#1}}%
}

\graphicspath{ {./images/} }

\newcommand\subfig[1]{\textcolor{red}{#1}}

\usepackage{xcolor}
\usepackage{hyperref} 

\hypersetup{
    allcolors=red,
    urlcolor= black,                
}

\usepackage{listings}
\usepackage{mathtools}
\definecolor{dkgreen}{rgb}{0,0.6,0}
\definecolor{gray}{rgb}{0.5,0.5,0.5}
\definecolor{mauve}{rgb}{0.58,0,0.82}

\lstdefinelanguage{myLang}
{
  morekeywords=[1]{subConstruct, 
                     DType,
                     DVector,
                     MType,
                     DSet,
                     DVar,
                     Trans,
                     Dim,
                     Met,
                     Weg,
                     Scope,
                     Input,
                     Output,
                     Range,
                     Update,
                     Status,
                     },
  morekeywords=[2]{ AccD_Comp_Dist,
                    AccD_Dist_Select,
                    AccD_Iter, 
                    AccD_Update
                    },
  morekeywords=[3]{int,
                   float,
                   status},
  sensitive=true, 
  morecomment=[l]{//}, 
  morecomment=[s]{/*}{*/}, 
  morestring=[b]", 
}
\usepackage{algorithm,algpseudocode}
\usepackage{amsmath}
\usepackage{amsfonts}
\usepackage{amssymb}
\usepackage{array,multirow,graphicx}
\def \hfillx {\hspace*{-\textwidth} \hfill}

\usepackage{tikz}
\usepackage{xcolor}

\usepackage[numbers,sort&compress]{natbib}

\newcommand*{\affaddr}[1]{#1}
\newcommand*{\affmark}[1][*]{\textsuperscript{#1}}
\newcommand*{\email}[1]{\texttt{#1}}

\begin{document}

\title{AccD: A Compiler-based Framework for Accelerating Distance-related Algorithms on CPU-FPGA Platforms}

\author{
Yuke Wang\affmark[1], Boyuan Feng\affmark[1], Gushu Li\affmark[2], Lei Deng\affmark[2], Yuan Xie\affmark[2], and Yufei Ding\affmark[1]\\
\affaddr{\affmark[1]Department of Computer Science}\\
\affaddr{\affmark[2]Department of Electrical and Computer Engineering}\\
\email{\affmark[1]\{yuke\_wang,boyuan,yufeiding\}@cs.ucsb.edu} \\ 
\email{\affmark[2]\{gushuli,leideng,yuanxie\}@ece.ucsb.edu}\\
\affaddr{University of California, Santa Barbara}
}

\maketitle

\begin{abstract}
    As a promising solution to boost the performance of distance-related algorithms (\textit{e.g.}, K-means and KNN), FPGA-based acceleration attracts lots of attention, but also comes with numerous challenges.
    In this work, we propose \textbf{AccD}, a compiler-based framework for accelerating distance-related algorithms on CPU-FPGA platforms. Specifically, AccD provides a Domain-specific Language to unify distance-related algorithms effectively, and an optimizing compiler to reconcile the benefits from both the algorithmic optimization on the CPU and the hardware acceleration on the FPGA.
    The output of AccD is a high-performance and power-efficient design that can be easily synthesized and deployed on mainstream CPU-FPGA platforms. Intensive experiments show that AccD designs achieve $31.42\times$ speedup and $99.63\times$ better energy efficiency on average over standard CPU-based implementations.
\end{abstract}
\section{Introduction} 
Distance-related algorithm  (\textit{e.g.}, K-means~\cite{LloydKMeans}, KNN~\cite{altman1992introduction}, and N-body Simulation~\cite{NBody-simulation}) plays a vital role in many domains, including machine learning, computational physics, etc. However, these algorithms often come with high computation complexity, leading to poor performance and limited applicability.
To improve their performance, FPGA-based acceleration gains lots of interests from both industry and research field, given its great performance and energy-efficiency. However, accelerating distance-related algorithms on FPGAs requires non-trivial efforts, including the hardware expertise, time and monetary cost. While existing works try to ease this process, they inevitably fall in short in one of the following aspects.
\textbf{Rely on problem-specific design and optimization while missing effective generalization.} There is no such unified abstraction to formalize the definition and optimization of distance algorithms systematically. Most of the previous hardware designs and optimizations~\cite{KMeansMicroarray,lin2012k,kdtreeKMeanscolorimage, KNNfpgahls} are heavily coded for a specific algorithm (\textit{e.g.}, K-means), which can not be shared with different distance-related algorithms. Moreover, these "hard-coded" strategies could also fail to catch up with the ever-changing upper-level algorithmic optimizations and the underlying hardware settings, which could result in a large cost of re-design and re-implementation during the design evolvement.

 \textbf{Lack of algorithm-hardware co-design.} Previous algorithmic~\cite{elkan2003using, ding2015yinyang} and hardware optimizations~\cite{lin2012k, kdtreeKMeanscolorimage, KMeansMicroarray, multicoreKMeans, KNNfpgahls} are usually applied separately instead of being combined collaboratively. Existing algorithmic optimizations, most of which are based on \textit{Triangle Inequality (TI)}~\cite{elkan2003using, ding2015yinyang, Topframework, chen2017sweet}, are crafted for sequential-based CPU. Despite removing a large number of distance computations, they also incur high computation irregularity and memory overhead. Therefore, directly applying these algorithmic optimizations to massively parallel platforms without taking appropriate hardware-aware adaption could lead to inferior performance.

\textbf{Count on FPGAs as the only source of acceleration.} 
Previous works~\cite{ParallelArchitecturesKNN, IPcoresKNN, ParameterizedKMeans, Lavenier00fpgaimplementation, KMeansMicroarray, KNNfpgahls} place the whole algorithm on the FPGA accelerator without considering the assists from the computing resource on the host CPU. As a result, their designs are usually limited by the on-chip memory and computing elements, and cannot fully exploit the power of the FPGA. Moreover, they miss the full performance benefits from the heterogeneous computing paradigm, such as using the CPU for complex logic and control operations while offloading the compute-intensive tasks to the FPGA.

\textbf{Lack of well-structured design workflow.} Previous works~\cite{ParallelArchitecturesKNN, ParameterizedKMeans, kdtreeKMeanscolorimage, lin2012k, KNNfpgahls} follow the traditional way of hardware implementation and require intensive user involvement in hardware design, implementation, and extra manual tuning process, which usually takes long development-to-validation cycles. Also, the problem-specific strategy leads to a case-by-case design process, which cannot be widely applied to handle different problem settings.
\begin{figure*}[t]
    \centering
    \includegraphics[width=1.6\columnwidth]{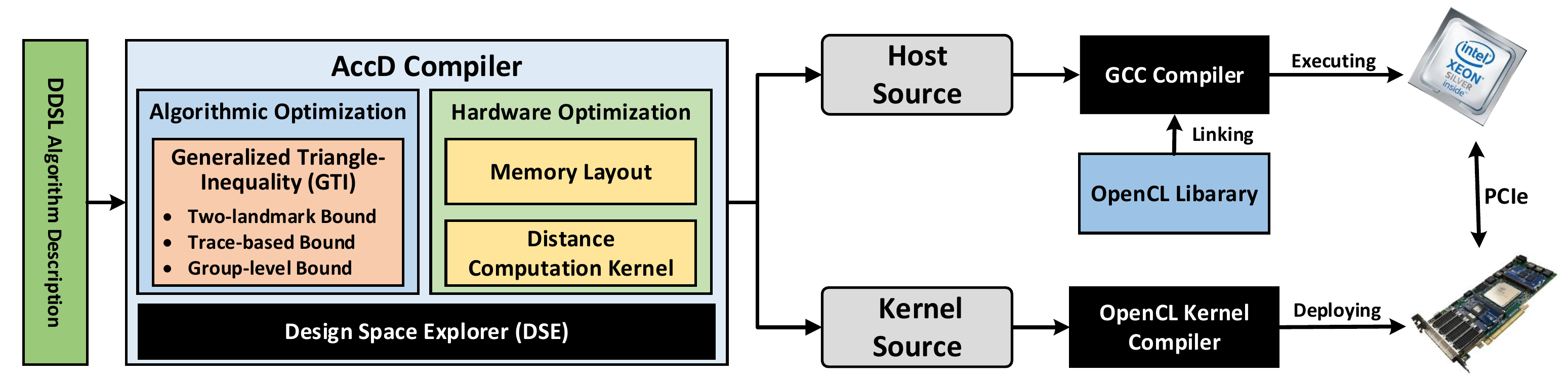}
    \caption{AccD Overview.}
    \label{fig: AccD Workflow}
\end{figure*}
To this end, we present a compiler-based optimization framework, \textit{AccD}, to automatically accelerate distance-related algorithms on the CPU-FPGA platform (shown in Figure~\ref{fig: AccD Workflow}). 
First, AccD provides a \textbf{Distance-related Domain-Specific Language} (\textit{DDSL}) as a problem-independent abstraction to unify the description and optimization of various distance-related algorithms. With the assist of the \textit{DDSL}, end-user can easily create highly-efficient CPU-FPGA designs by only focusing on high-level problem specification without touching the algorithmic optimization or hardware implementation.   

Second, AccD offers a novel \textbf{algorithmic-hardware co-optimization} scheme to reconcile the acceleration from both sides. At the algorithmic level, AccD incorporates a novel \textit{Generalized Triangle Inequality (GTI)} optimization to eliminate unnecessary distance computations, while maintaining the computation regularity to a large extent. At the hardware level, AccD employs a \textit{specialized data layout} to enforce memory coalescing and an \textit{optimized distance computation kernel} to accelerate the distance computations on the FPGA. 

Third, AccD leverages both the host and accelerator side of the \textbf{CPU-FPGA heterogeneous system for acceleration}. In particular, AccD distributes the algorithm-level optimization (\textit{e.g.}, data grouping and distance computation filtering) to CPU, which consists of complex operations and execution dependency, but lacks pipeline and parallelism. On the other hand, AccD assigns hardware-level acceleration (\textit{e.g.}, distance computations) to the FPGA, which is composed of simple and vectorizable operations. Such mapping successfully capitalizes the benefit of CPU for managing control-intensive tasks and the advantage of FPGA for accelerating computation-intensive workloads. 
Lastly, AccD compiler integrates an intelligent \textbf{Design Space Explorer} \textit{(DSE)} to pinpoint the "optimal" design for different problem settings. In general, there is no existing "one size fits all" solution: the best configuration for algorithmic and hardware optimization would differ across different distance-related algorithms or different inputs of the same distance-related algorithm. 
To produce a high-quality optimization configuration automatically and efficiently, DSE combines the design modeling (performance and resource) and Genetic Algorithm to facilitate the design space search.

Overall, our contributions are:
\begin{itemize}    
    \item We propose the first optimization framework that can automatically optimize and generate high-performance and power-efficient designs of distance-related algorithms on CPU-FPGA heterogeneous computing platforms.  
    \item We develop a Domain-specific Language, DDSL, to unify different distance-related algorithms in an effective and succinct manner, laying the foundation for general optimizations across different problems.
    \item We build an optimizing compiler for the DDSL, which automatically reconciles the benefits from both the algorithmic optimization on CPU and hardware acceleration on FPGA.
    \item Intensive experiments on several popular algorithms across a wide spectrum of datasets show that AccD-generated CPU-FPGA designs could achieve $31.42\times$ speedup and $99.63\times$ better energy-efficiency on average compared with standard CPU-based implementations.
\end{itemize}
\section{Related Work}      \label{sect: Related Work}
\vspace*{-0.32em}
Previous research accelerates distance-related algorithms in two aspects: \textit{Algorithmic Optimization} and \textit{Hardware Acceleration}. More details are discussed in the following subsections.

\subsection{Algorithmic Optimization} 
From the algorithmic standpoint, previous research highlights two optimizations. The first one is KD-tree based optimization~\cite{KD-TreeKMeans, efficientKmeans, KNNJoinsDataStreams, 5952342, Zhong:2013:GEI:2505515.2505749}, which relies on storing points in special data structures to enable nearest neighbor search without computing distances to all target points. These methods often deliver $3\times \sim 6\times$ performance improvement~\cite{KD-TreeKMeans, efficientKmeans, KNNJoinsDataStreams, 5952342, Zhong:2013:GEI:2505515.2505749} compared with the unoptimized versions in low dimensional space, while suffering from a serious performance degradation when handling large datasets with high dimension ($d \geq 20$) due to their exponentially-increased memory and computation overhead.

The second one is TI based optimization~\cite{elkan2003using, ding2015yinyang, Topframework, chen2017sweet}, which aims at replacing computation-expensive distance computations with cheaper bound computations, demonstrates its flexibility and scalability. It can not only reduce the computation complexity at different levels of granularity but is also more adaptive and robust to the datasets with a wide range of size and dimension. 
However, most existing works focus on one specific algorithm (\textit{e.g.}, KNN~\cite{chen2017sweet}, K-means~\cite{elkan2003using, ding2015yinyang}, etc.), which lack extensibility and generality across different distance-related problems. 
An exception is a recent work, TOP~\cite{Topframework}, which builds a unified framework to optimize various distance-related problems with pure TI optimization on CPUs. Our work shares a similar high-level motivation with their work, but targets at a more challenging scenario: algorithmic and hardware co-optimization on CPU-FPGA platforms. 

\subsection{Hardware Acceleration}
From the hardware perspective, several FPGA accelerator designs have been proposed, but still suffer from some major limitations.

First, previous FPGA designs are generally built for specific distance-related algorithm and hardware. For example, works from~\cite{KMeansMicroarray, kdtreeKMeanscolorimage, lin2012k} target on KNN FPGA acceleration, while researches from~\cite{KNNfpgahls, IPcoresKNN, ParallelArchitecturesKNN} focus on K-means. Moreover, previous designs~\cite{lin2012k, KMeansMicroarray} usually assume that dataset can be fully fit into the FPGA on-chip memory, and they are only evaluated on a limited number of small datasets, for example, in~\cite{lin2012k}, K-means acceleration is evaluated on a micro-array dataset with only 2,905 points. These designs often encounter portability issues when transferring to different settings. Besides, these "hard-coded" designs and optimizations create difficulties for a fair comparison among different designs, which hamper future studies in this direction. 

The second problem with previous works is that they fail to incorporate algorithmic optimizations in the hardware design.
For example, works from~\cite{KMeansMicroarray, ParallelArchitecturesKNN, kdtreeKMeanscolorimage, KNNfpgahls}, directly port the standard K-means and KNN algorithms to FPGA, and only apply hardware-level optimization. 
One exception is a recent work~\cite{KPynq}, which promotes to combine TI optimization and FPGA acceleration for K-means. It gives a considerable speedup compared to state-of-the-art methods, showcasing the great opportunity of applying algorithm-hardware co-optimization. Nevertheless, this idea is far from well-explored, possibly because it requires the domain knowledge and expertise from both the algorithm and hardware to combine both of them effectively. 

In addition, previous works largely focus on the traditional hardware design flow, which requires a long implementation cycle and huge manual efforts. For example, works from~\cite{ParameterizedKMeans, KMeansMicroarray, multicoreKMeans, kdtreeKMeanscolorimage, ICSICT2016, Lavenier00fpgaimplementation, adaptiveKNNPartialReconfiguration, adaptiveKNN} build the design based on VHDL/Verilog design flow, which requires hardware expertise and over months of arduous development. In contrast, our AccD design flow brings significant advantages of programmability and  flexibility due to its high-level OpenCL-based programming model, which minimizes the user involvement in the tedious hardware design process.
\section{\textbf{D}istance-related Algorithm \textbf{D}omain-\textbf{S}pecific \textbf{L}anguage (DDSL)} 
\label{sect: DDSL}
Distance-related algorithms share commonalities across different application domains and scenarios, even though they look different in their high-level algorithmic description. Therefore, it is possible to generalize these distance-related algorithms. AccD framework defines a DDSL, which provides a high-level programming interface to describe distance-related algorithms in a unified manner. Unlike the API-based programming interface used in the TOP framework~\cite{Topframework}, DDSL is built on C-like language and provides more flexibility in low-level control and performance tuning support, which is crucial for FPGA accelerator design.

Specifically, DDSL utilizes several constructs to describe the basic components (\textbf{Definition}, \textbf{Operation}, and \textbf{Control}) of the distance-related algorithms, and also identify the potential parallelism and pipeline opportunities during the design time. We detail these constructs in the following part of this section. 

\subsection{Data Construct}
Data construct is a basic \textbf{Definition Construct}. It leverages \code{DSet} primitive to indicate the name of the data variable, and the \code{DType} primitive to notate the type characters of the defined variable. Data construct serves as the basis for AccD compiler to understand the algorithm description input, such as the data points that are used in the distance-related algorithms. An example of data constructs is shown in the code below, where we define the variable and dataset using DDSL data construct.
\vspace*{-0.3\baselineskip}
\begin{lstlisting}
/* Define a single variable */
DVar [setName] DType [Optional_Initial_Value];
/* Define the matrix of dataset */
DSet [setName] DType [size] [dim];
\end{lstlisting}
\vspace*{-0.3\baselineskip}

In most distance-related algorithms, the dataset can be defined as the source set and the target set. For example, in K-means, the source set is the set of data points, and the target set is the set of clusters. Currently, AccD supports several data types including \textit{int} (32-bit), \textit{float} (32-bit), \textit{double} (64-bit) based on the users' requests, algorithm performance, and accuracy trade-offs.

\subsection{Distance Compute Construct}
Distance computation is the core \textbf{Operation Construct} for distance-related algorithms, which measures the exact distance between two different data points. This construct requires several fields, including data dimensionality, distance metrics, and weight matrix (if weighted distance is specified).
\begin{lstlisting}
AccD_Comp_Dist(Input p1, Input p2, Output disMat, Output idMat, Dim dim, Met mtr, Weg mat)
\end{lstlisting}
\vspace{-1.5em}
\begin{table}[ht] \small
\centering
 \begin{tabular}{|| c | l ||} 
 \hline
     p1, p2 & Input data matrix. ($n_1 \times d$, $n_2 \times d$)\\ 
     \hline
     disMat &  Output distance matrix. ($n_1 \times n_2$)\\
     \hline
     idMat & Output id matrix. ($n_1 \times n_2$) \\
     \hline
     dim & Dimensionality of input data point.\\
     \hline
     mtr &  Distance metric:\code{(Weighted|Unweighted)}\\ \hline
     mat & Weight matrix: Used for weighted distance ($1\times d$)\\
    \hline
\end{tabular}
\caption{Distance Compute Construct Parameters.}
\label{fig: Distance Compute Construct Parameters.}
\end{table}

\vspace*{-1.3\baselineskip}
\subsection{Distance Selection Construct}
Distance selection construct is an \textbf{Operation Construct} for distance value selection and it returns the Top-K smallest or largest distances and their corresponding points ID number from the provided distance and ID list. This construct helps AccD compiler to understand the distances of users' interests.
\begin{lstlisting}
AccD_Dist_Select(Input distMat, Input idMat, Output TopKMat, Range ran, Scope scp)
\end{lstlisting}
\begin{table}[h]  \small
\centering
 \begin{tabular}{|| c | p{18em}||} 
 \hline
     TopKMat   &  Top-K id matrix ($n_1 \times k$)\\ 
     \hline
     ran      & Scalar value of \code{K} (\textit{e.g.}, K-means, KNN) or distance threshold (\textit{e.g.}, N-body Simulation)\\
     \hline
     scp     & Top-K \code{(smallest|largest)} values\\
 \hline
\end{tabular}
\caption{Distance Selection Construct Parameters.}
\label{fig: Distance Selection Construct Parameters.}
\end{table}
\vspace*{-1.1\baselineskip}
\subsection{Data Update Construct}
\vspace*{-0.2\baselineskip}
Data update construct is an \textbf{Operation Construct} for updating the data points based on the results from the prior constructs. For example, K-means updates the cluster centers by averaging the positions of the points inside. This construct requires the variable to be updated and additional information to finish this update, such as the point-to-cluster distances. The status of this data update will be returned after the completion of all its inside operations. The status variable is to tell whether the data update makes a difference or not.
\begin{lstlisting} 
AccD_Update(Update var, Input p1 ,..., Input pm, Status s)
\end{lstlisting}
\vspace*{-1.2\baselineskip}
\begin{table}[ht]  \small
\centering
 \begin{tabular}{|| c | l ||} 
 \hline
     upVar & Input data/dataset to be updated \\
     \hline
     {p1, ..., pm} & Additional information used in update\\ 
     \hline
     $S$ & Status of update operation.\\
 \hline
\end{tabular}
\caption{Data Update Construct Parameters.}
\label{fig: Data Update Construct Parameters.}
\end{table}

\vspace*{-1.5\baselineskip}
\subsection{Iteration Construct}
\vspace*{-0.5\baselineskip}
Iteration construct is a top-level \textbf{Control Construct}. It is used to describe the distance-related algorithms that require iteration, such as K-means. Iteration construct requires users to provide either the maximum number of iteration or other exit condition.
\begin{lstlisting}[mathescape=true]
AccD_Iter(maxIterNum|exitCond){
    subConstruct $sc_1$;
    subConstruct $sc_2$;
    ...
    subConstruct $sc_n$;
}
\end{lstlisting}

\vspace*{-0.5\baselineskip}
\subsection{Example: K-means}
\vspace*{-0.2em}
To show the expressiveness of DDSL, we take K-means as an example. From the code shown below, with no more than 20 lines of code, DDSL can capture the key components of user-defined K-means algorithm, which is essential for AccD compiler to generate designs for CPU-FPGA platforms.
\vspace{0.2em}
\begin{lstlisting}
DVar K int 10;
DVar D int 20;
DVar psize int 1400;
DVar csize int 200;
DSet pSet float psize D;
DSet cSet float csize D;
DSet distMat float psize csize;
DSet idMat int psize csize;
DSet pkMat int psize K;
AccD_Iter(S){
    S = false;
    /* Compute the inter-dataset distances */
    AccD_Comp_Dist(pSet, cSet, distMat, idMat, D, "Unweighted L1", 0);
    /* Select the distances of interests */
    AccD_Dist_Select(distMat, idMat, K, "smallest", pkMat);
    /* Update the cluster center */
    AccD_Update(cSet, pSet, pkMat, S)
}
\end{lstlisting}
\section{Algorithm Optimization} \label{sect: Algorithm Optimization}
This section explains a novel TI optimizations tailored for CPU-FPGA platforms. TI has been used for optimizing distance-related problems, but is often on the sequential processing systems. Our design features an innovative way of applying TI to obtain low-overhead distance bounds for unnecessary distance computation elimination while maintaining the computation regularity to ease the hardware acceleration on FPGAs.

\subsection{TI in Distance-related Algorithm}
\begin{figure*}[htb] \small
    \centering
    \makebox{\includegraphics[scale=0.8]{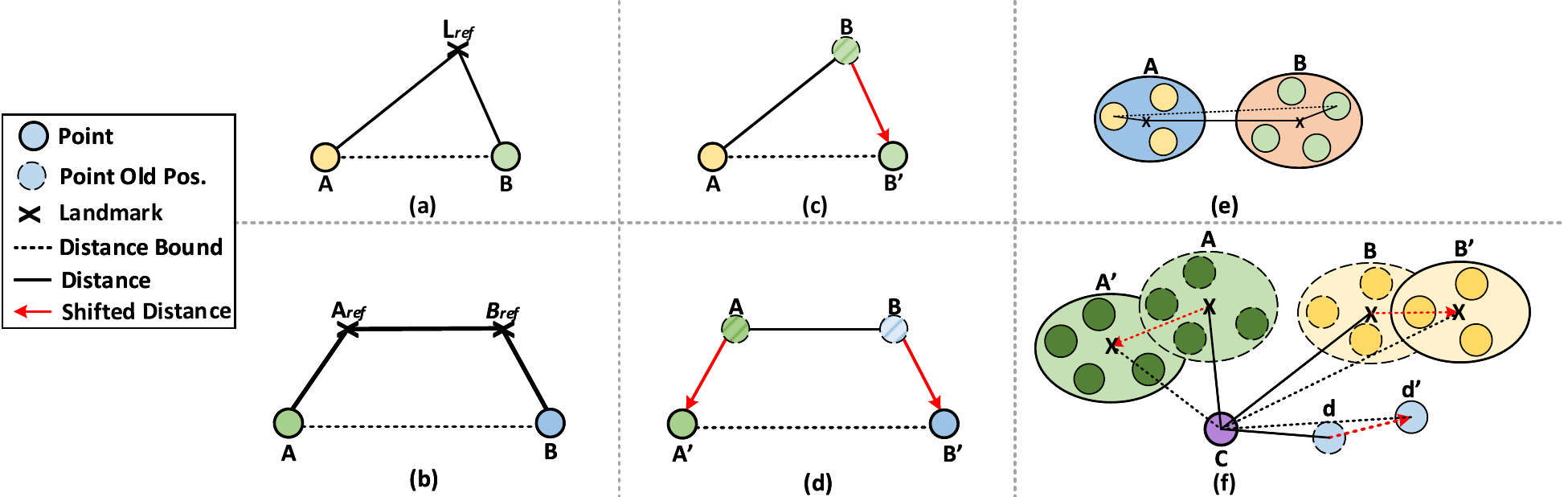}}
    \caption{TI Optimization.}
    \label{fig: TI Optimization.}
\end{figure*}

As a simple but powerful mathematical concept, TI has been used to optimize the distance-related algorithm.  Figure~\ref{fig: TI Optimization.}\subfig{a} gives an illustration. It states that $d(A, B) \leq d(A, L_{ref}) + d( L_{ref}, B)$, where $d(A, B)$ represents the distance between point $A$ and $B$ in some metrics (\textit{e.g.}, Euclidean distance). The assistant point $L_{ref}$ is a landmark point for reference. Directly from the definition, we could compute both the lower bound ($lb(A, B)$) and upper bound ($ub(A, B)$) of the distance between two points A and B. This is the standard and most common usage of TI for deriving bounds of distance.

In general, bounds can be used as a substitute for the exact distances in the distance-related data analysis. Take N-body simulation as an example. It requires to find target points that are within $R$ (the radius) from each given query point. Suppose we get the $lb(A, B) = 10$ and $10 > R$, then we are 100\% confident that source point $A$ is not within $R$ of query point $B$. As a result, there is no need to compute the exact distance between point $A$ and $B$. Otherwise, the exact distance computation will still be carried out for direct comparison. 
While many previous researches~\cite{elkan2003using, ding2015yinyang, lin2012k, MakingKMeansFaster, KNNAdaptiveBound} gain success in directly porting the above point-based TI to optimize distance-related algorithms, they usually suffer from memory overhead and computations irregularity, which result in inferior performance. 
\vspace{-1em}
\subsection{Generalized Triangle Inequality (GTI)} 
AccD uses a novel Generalized TI (GTI) to remove redundant distance computation. It generalizes the traditional point-based TI while significantly reducing the overhead of bound computations. The traditional point-based TI focuses on tighter bound (more closer to the exact distance) to remove more distance computations, but it induces the extra bound computations, which could become the new performance bottleneck even after many distance calculations being removed. In contrast, GTI strikes a good balance between distance computation elimination and bound computation overhead. In particular, AccD highlights GTI from three perspectives: \textit{Two-landmark bound computation}, \textit{Trace-based bound computation}, and \textit{Group-level bound computation}.

\paragraph{Two-landmark Bound Computation} 
Two-landmark scheme aims at reducing the bound computation through effective distance reuse. In this case, the distance bound between two points can be measured through two landmarks as the reference points. As illustrated in Figure \ref{fig: TI Optimization.}\subfig{b}, the distance bound between point $A$ and $B$ can be computed based on $d(A, A_{ref})$, $d(B, B_{ref})$ and $d(A_{ref}, B_{ref})$ through Equation \ref{equ: Two-landmark Bound Computation.}, where $A_{ref}$ and $B_{ref}$ are the landmark points for point $A$ and $B$, correspondingly. 
\begin{equation} \small
\label{equ: Two-landmark Bound Computation.}
\begin{aligned}
    lb(A, B) \geq  d(A_{ref}, B_{ref}) - d(A, A_{ref}) - d(B, B_{ref}) \\
    ub(A, B) \leq  d(A_{ref}, B_{ref}) + d(A, A_{ref}) + d(B, B_{ref})
\end{aligned}
\end{equation}

One representative application scenario of Two-landmark bound computation is KNN-join, where two disjoint sets of landmarks are selected for the query and target point set. In this case, much fewer bound computations are required compared with the one-landmark case (shown in Figure~\ref{fig: TI Optimization.}\subfig{a}). This can also be validated through a simple calculation. Assuming in KNN-join, we have $m$ query points, $n$ target points, $z_{qry}$ query landmarks, and $z_{trg}$ target landmarks. Also,  we have $z_{qry}<<m$ and $z_{trg}<<n$ in general. Therefore, we can get $m + n + z_{qry} \times z_{trg}$ bound computations for Two-landmark case, which is much smaller than One-landmark bound computation ($m \times z_{trg} + n$ or $n \times z_{qry} + m$).

\paragraph{Trace-based Bound Computation} 
Trace-based bound computation finds its strength in iterative distance algorithms with points update, since it can largely reduce the bound computation overhead over numbers of iterations. The key to Trace-based bound computation is selecting appropriate landmark points as references. For example, in K-means, only the target points (clusters) change their positions across iterations, therefore, we can choose the previous positions of clusters from the last iteration as the landmarks for bound computation in the current iteration, since these "old" cluster positions
can be close enough to the current point positions to offer "tight" bound. This process can be illustrated in Figure~\ref{fig: TI Optimization.}\subfig{c}, where the distance bound $d(A, B')$ can be calculated based on $d(B, B')$ and $d(A, B)$, where $B'$ is the new point position while $B$ is the old point position from the last iteration. 

In addition, Trace-based bound computation can also work collaboratively with the Two-landmark cases. For example, in N-body simulation, the source and target points are essentially the same dataset and would get updated across iterations. We can choose the "old" position of each point from the last iteration as the landmark for the bound computation at the current iteration, due to its closeness towards the current point position. 
This case can be clarified in Figure~\ref{fig: TI Optimization.}\subfig{d}, where $A$ and $B$ are the "old" point positions from the last iteration, $A'$ and $B'$ are the new source and target point. Then based on $d(A, B)$, $d(A, A')$ and $d(B, B')$, the new distance bound between $A'$ and $B'$ can be easily derived by using the old points $A$ and $B$ as the reference points. And the cost of this is also as low as $O(n)$, where $n$ is the number of particles, since each point only need to maintain the shifted distance between its new position and old position from the last iteration (Note: we only compute the inter-point distances at the first iteration). In contrast, only applying Two-landmark without the effective temporal reuse of the old point position will result in the complexity at least $O(n \times z)$, where $z$ is number of landmarks. Since, in this case, distances between each point and all the landmarks have to be computed, so that each point can know its new closest landmark before applying the bound computation. 

\paragraph{Group-level Bound Computations} 
Group-level bound computation aims at reducing the bound computation overhead while maintaining the computation regularity. Group-level bound computations features itself with the capability to combine with aforementioned two bound cases as the hybrid bound computation. In the combination with the Two-landmark case, as shown in Figure \ref{fig: TI Optimization.}\subfig{e}, points in each group ($A$ and $B$) share the same landmark ($A_{ref}$ and $B_{ref}$) as the reference point. Then based on $d(A_{ref}, B_{ref})$ and the $d_{max}(a, A_{ref})$ and $d_{max}(b, B_{ref})$, we can get the group-level bound based on Equation \ref{equ: Group-level + Two-landmark Bound Computation.}, where $d_{max}(a, A_{ref})$ and $d_{max}(b, B_{ref})$ get the distance between the farthest point within each group and its group reference point.  
\begin{equation} \small
\label{equ: Group-level + Two-landmark Bound Computation.}
\begin{aligned}
    lb(A, B) \geq  d(A_{ref}, B_{ref}) - d_{max}(a, A_{ref}) - d_{max}(b, B_{ref}) \\
    ub(A, B) \leq  d(A_{ref}, B_{ref}) + d_{max}(a, A_{ref}) + d_{max}(b, B_{ref})
\end{aligned}
\end{equation}
\vspace{-0.7em}

In the combination with the Trace-based case, it will generate a hierarchical bound as a hybrid solution, which includes point-group bound and point-point bound computation. As exemplified in Figure \ref{fig: TI Optimization.}\subfig{f}, each group regards its old group center as the landmark for reference, and each point relies on its old position as the landmark for reference. Then based on $d(A, A')$, $d(B, B')$, and $d(d, d')$, and the old distance $d(c, A)$, $d(c,B)$ and $d(c, d)$, where $A$ and $B$ are the point groups, and point $d$ is the closest point of point $c$ in the last iteration. We can calculate $lb(c, G')$ and $ub(c, d')$ based on Equation \ref{equ: hierarchy bound computation.}, 
\begin{equation} \small
\label{equ: hierarchy bound computation.}
\begin{aligned}
    lb(c, G') \geq d(c, G) - d_{max}(G, G') \\ 
    ub(c, d') \leq d(c, d) + d_{max}(d, d') \\ 
\end{aligned}
\end{equation}
where $G \in \{A, B\}$. If we have $d(c, G) - d_{max}(G, G') > d(c, d) + d_{max}(d, d')$, it is impossible that the points inside the group $A'$ and $B'$ can become the closest point of $c$ in the current iteration. Therefore, the distance computation between the point $c$ and all points inside these groups can be safely avoided.

\begin{figure} [ht] \small
    \centering
    \makebox{\includegraphics[width=0.75\columnwidth]{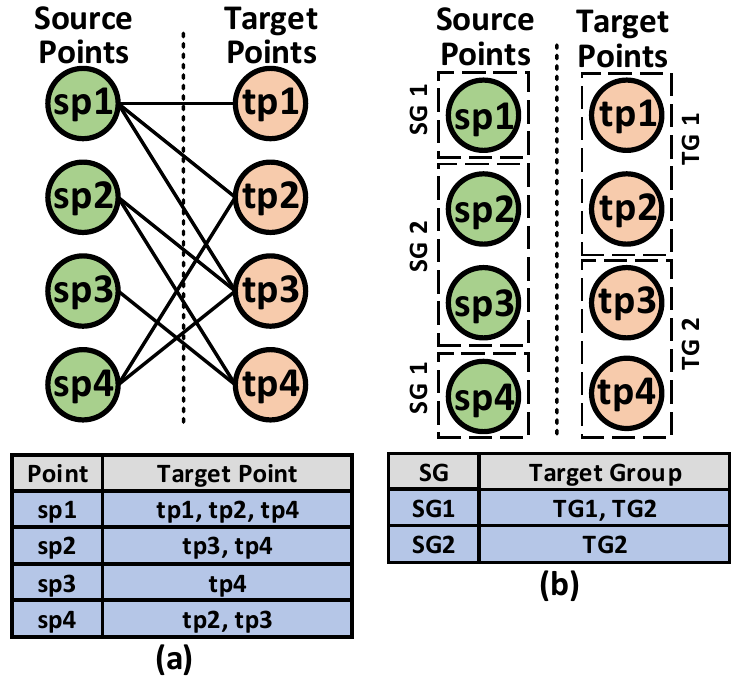}}
    \caption{Bound Computation at (a) Point-level, (b) Group-level.}
    \label{fig: group-level bound computation -- computation regularity.}
\end{figure}

In addition to distance saving, group-level bound computation offers another two benefits to facilitate the underlying hardware acceleration. First, the computation regularity on the remaining distance computation becomes higher compared with the point-level bound computation. Since points inside each group will share the commonality in computation, which facilitates the parallelization for acceleration. For example, point-level bound computation usually results in a large divergence of distance computation among different points, as shown in Figure~\ref{fig: group-level bound computation -- computation regularity.}\subfig{a}, which is a killer of parallelization and pipelining. However, in group-level bound computation, points inside the same source group will always maintain the same groups of target points for distance computation, as shown in Figure~\ref{fig: group-level bound computation -- computation regularity.}\subfig{b}. 
\vspace*{-0.8em}
Second, group-level bound computation brings the benefit of reducing memory overhead. Assuming we have $m$ source points, $n$ target points, $z_{src}$ source groups, and $z_{trg}$ target groups. The memory overhead of maintaining distance bounds is $O(m\times n)$ in the point-level bound computation case. However, in the group-level bound computation case, we only have to maintain distance bounds among groups, and the memory overhead is $O(z_{src}\times z_{trg})$, where $z_{src} << m$ and $z_{trg}<< n$. Therefore, in terms of memory efficiency, group-level bound computation can outperform the point-level bound computation to a great extent.

\section{Hardware Acceleration} \label{sect: Architecture Design}
AccD design is built on the CPU-FPGA architecture, which highlights its significant performance and energy efficiency, and has been widely adopted as the modern data center solution for high-performance computing and acceleration. The host-side application of AccD design is responsible for data grouping and distance computation filtering, which consists of complex operations and execution dependency, but lacks pipeline and parallelism. On the other hand, the FPGA-side of AccD design is built for accelerating the distance computations, which are composed of simple and vectorizable operations.

While FPGA accelerator features with high computation capability, the memory bandwidth bottleneck constraints the overall design performance. Therefore, optimizing data placement and memory architecture is the key to improving memory performance. In addition, the OpenCL-based programming model adds a layer of architectural complexity of the kernel design and management, which is also critical to the design performance. AccD framework distinguishes itself by using a novel memory and kernel optimization strategy that is tailored for TI-optimized distance-related algorithms to benefit CPU-FPGA designs.

\subsection{Memory Optimization}
After applying the GTI optimization to remove the redundant distance computation, each source point group will have different target groups as candidates for distance computation, as shown in Figure~\ref{table: inter-group memory.}\subfig{a}, where \textbf{Source-grp} is ID of the source group, and \textbf{Target-grp} is ID of the target group. However, this would raise two concerns about performance degradation. 
\begin{table}[ht] \small
\begin{minipage}[b]{0.5\columnwidth}
\centering
\begin{tabular}{|| c | c ||}
\hline
\textbf{Source-grp}
& \makecell{\ \textbf{Target-grp}\\} 
\\ 
\hline
\hline  $s_1$  &	$t_1$, $t_4$, $t_6$
\\ 
\hline  $s_2$  &	$t_8$, $t_{10}$, $t_{12}$
\\
\hline  ...  &	...
\\
\hline  $s_5$  &   $t_2$, $t_4$, $t_6$
\\
\hline  $s_6$  &  $t_8$, $t_{10}$, $t_{12}$
\\
\hline
\end{tabular}
\caption*{(a)}
\end{minipage}
\hfillx
\begin{minipage}[b]{0.5\columnwidth}
\centering
\label{table: Optimized Memory}
\begin{tabular}{|| c | c ||}
\hline
\textbf{Source-grp}
& \makecell{\ \textbf{Target-grp}\\} 
\\ 
\hline
\hline  $s_1$  &	$t_2$, $t_4$, $t_6$
\\ 
\hline  $s_5$  &	$t_2$, $t_4$, $t_6$
\\
\hline  $s_2$  &	$t_8$, $t_{10}$, $t_{12}$
\\
\hline  $s_6$  &   $t_8$, $t_{10}$, $t_{12}$
\\
\hline  ...  &	...
\\
\hline
\end{tabular}
\caption*{(b)}
\end{minipage}
\captionof{figure}{(a) Non-optimized inter-group memory access; (b) Optimized inter-group memory access.}
\label{table: inter-group memory.}
\end{table}
\vspace{-0.8em}

The first issue is inter-group memory irregularity and low data reuse. For example, the target group information ($t_1$, $t_4$, $t_6$) required by source group $s_1$ can not be reused by $s_2$. Since $s_2$ requires quite different target groups ($t_8$, $t_{10}$, and $t_{12}$) for distance computation, thus, additional costly memory access has to be carried out. To tackle this problem, AccD places the source groups to the continuous memory space to maximize the memory access efficiency, only if these source groups have the same set of target groups as candidates for distance computation. An example has been shown in Figure~\ref{table: inter-group memory.}\subfig{b}, where the source group $s_2$ and $s_6$ are placed side by side in the memory, since they have the same list of target groups ($t_8$, $t_{10}$, and $t_{12}$), which can take advantage of the memory temporal locality without issuing another memory access.

\begin{figure}[ht]
\begin{minipage}[c]{0.3\columnwidth}
\centering
\begin{tabular}{|| c | c ||}
\hline
\textbf{Group}
& \makecell{\ \textbf{Points}\\} 
\\ 
\hline
\hline  $Grp_1$  & $3$, $8$, $9$
\\ 
\hline  $Grp_2$  &	$5$, $6$, $7$ 
\\
\hline  $Grp_3$ &	$1$, $2$, $4$
\\
\hline  ...  &	...
\\
\hline
\end{tabular}
\caption*{(a)}
\label{table: Group Point Mapping}
\end{minipage}
\qquad
\begin{minipage}[c]{0.3\columnwidth}
    \centering
    \includegraphics[width=0.7\columnwidth]{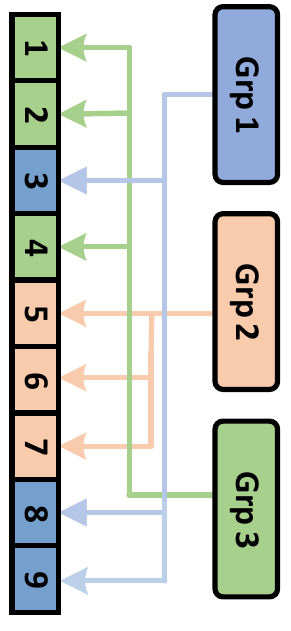}
    \caption*{(b)}
\end{minipage} 
\hfill 
\begin{minipage}[c]{0.3\columnwidth}
    \centering
    \includegraphics[width=0.92\columnwidth]{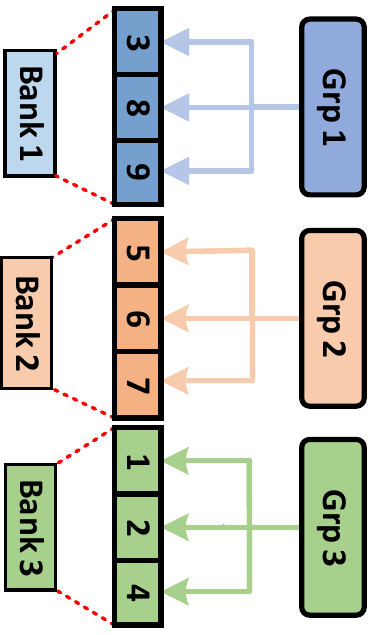}
    \caption*{(c)}
\end{minipage}
\captionof{figure}{(a) Group-point mapping; (b) Non-aligned intra-group memory; (c) Aligned intra-group memory.}
\label{fig: memory optimization.}
\end{figure}

The second issue is intra-group memory irregularity. For example, points from \textit{group 1}, \textit{2}, and \textit{3} have taken up the memory space at intervals, as shown in Figure~\ref{fig: memory optimization.}\subfig{a}. However, a group of points are usually accessed simultaneously due to GTI optimization. This would cause frequent inefficient memory access for fetching individual point distributed at the discontinuous memory address. To solve this issue, AccD offers a second memory optimization to re-organize the target/source points inside the same target/source group into continuous memory space within the same memory bank, as illustrated in Figure~\ref{fig: memory optimization.}\subfig{b}. This strategy can largely benefit memory coalescing and external memory bandwidth while minimizing the access contention, since points inside the same bank can be accessed efficiently and points inside different banks can be accessed in parallel. 
\vspace*{-1em}
\subsection{Distance Computation Kernel}
Distance computation takes the major time complexity in distance-related algorithms. In AccD, after TI filtering on CPU, the remaining distance computations are accelerated on FPGA. Points involved in the remaining distance computations are organized into two sets: source set and target set, which can be organized as two matrices, $Mat_{A}$ ($m\times d$) and $Mat_{B}$ ($n\times d$), respectively, where each row of these matrices represents a point with $d$ dimension. The distance computation between $Mat_A$ and $Mat_B$ can be decomposed into three parts, as shown in Equation~\ref{equ: distance computation},
\begin{equation} \small
    \label{equ: distance computation}
    {(Mat_{A} - Mat_{B})}^2 = Mat_{A}^2 - 2 * Mat_{A} \cdot Mat_{B} + Mat_{B}^2
\end{equation}
where $Mat_{A}^2$ or $Mat_{B}^2$ only takes the complexity of $O(m\times d)$ and $O(n\times d)$, while $Mat_A\times Mat_ B$ takes $O(m\times n\times d^2)$, which dominates the overall computation complexity. AccD spots an efficient way of accelerating $Mat_A\cdot Mat_B$ through highly-efficient matrix-matrix multiplication, which can benefit the hardware implementation on FPGA.
\begin{figure} [ht!] \small
    \centering
    \includegraphics[width=0.8\columnwidth]{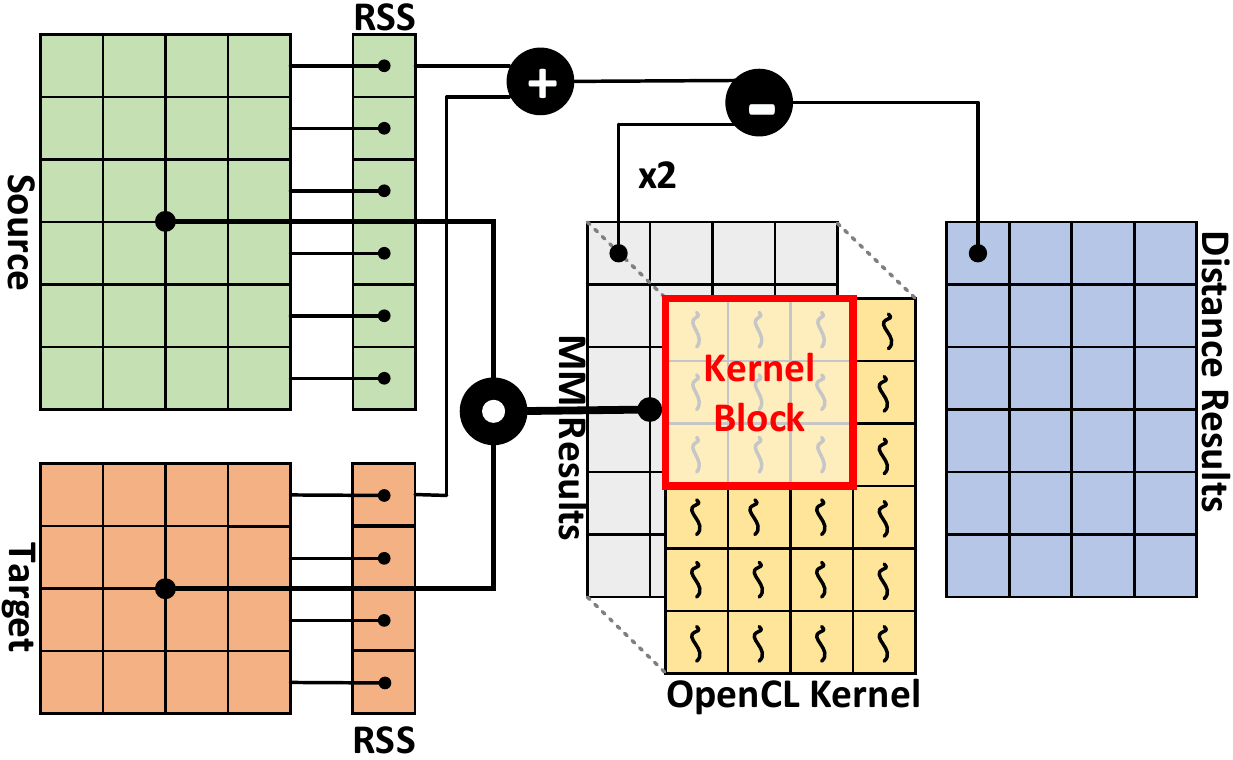}
    \caption{AccD Matrix-based Distance Computation.}
    \label{fig: AccD Distance Computation.}
\end{figure}

The overall computation process can be described as Figure~\ref{fig: AccD Distance Computation.}, the source ($Mat_A$) and target set ($Mat_B$) Row-wise Square Sum (RSS) is pre-computed through in a fully-parallel manner. And the vector multiplication between each source and target point is mapped to an OpenCL kernel thread for a fine-grained parallelization. Moreover, a block of threads, as highlighted in the "red" square box of Figure~\ref{fig: AccD Distance Computation.}, is the kernel thread workgroup, which can share a part of the source and target points to increase the on-chip data locality. Based on this kernel organization, AccD hardware architectural design offers several tunable hyperparameters for performance and resource trade-off: the size of kernel block, the number of parallel pipeline in each kernel block, etc. To efficiently find the "optimal" parameters that can maximize overall performance while respecting the constraints, we harness the AccD explorer for efficient design space search, which is detailed in Section~\ref{sect: Design Space Exploration}.
\section{AccD Compiler} \label{sect: AccD Compiler}
In this section, we detail AccD compiler in two aspects: design parameters and constraints, and design space exploration.
\vspace{-0.2em}
\subsection{Design Parameters and Constraints}
AccD uses a parameterized design strategy for better design flexibility and efficiency. It takes the design parameters and constraints from algorithm and hardware to explore and locate the "optimal" design point tailored for the specific application scenario. At the algorithm level, the number of groups affects distance computation filtering performance. At the hardware level, there are three parameters: 1) Size of computation block, which decides the size of data shared by a group of computing elements; 2) SIMD factor, which decides the number of computing elements inside each computation block; 3) Unroll factor, which tells the degree of parallelization in each single distance computation. In addition, there are several hardware constraints, such as the on-chip memory size, the number of logic units, and the number of registers. All of these parameters and constraints are included in our analytical model for design exploration.

\vspace{0.4em}
\subsection{Design Space Exploration} 
\label{sect: Design Space Exploration}
\begin{figure} [h] \small
    \centering
    \includegraphics[height=10em]{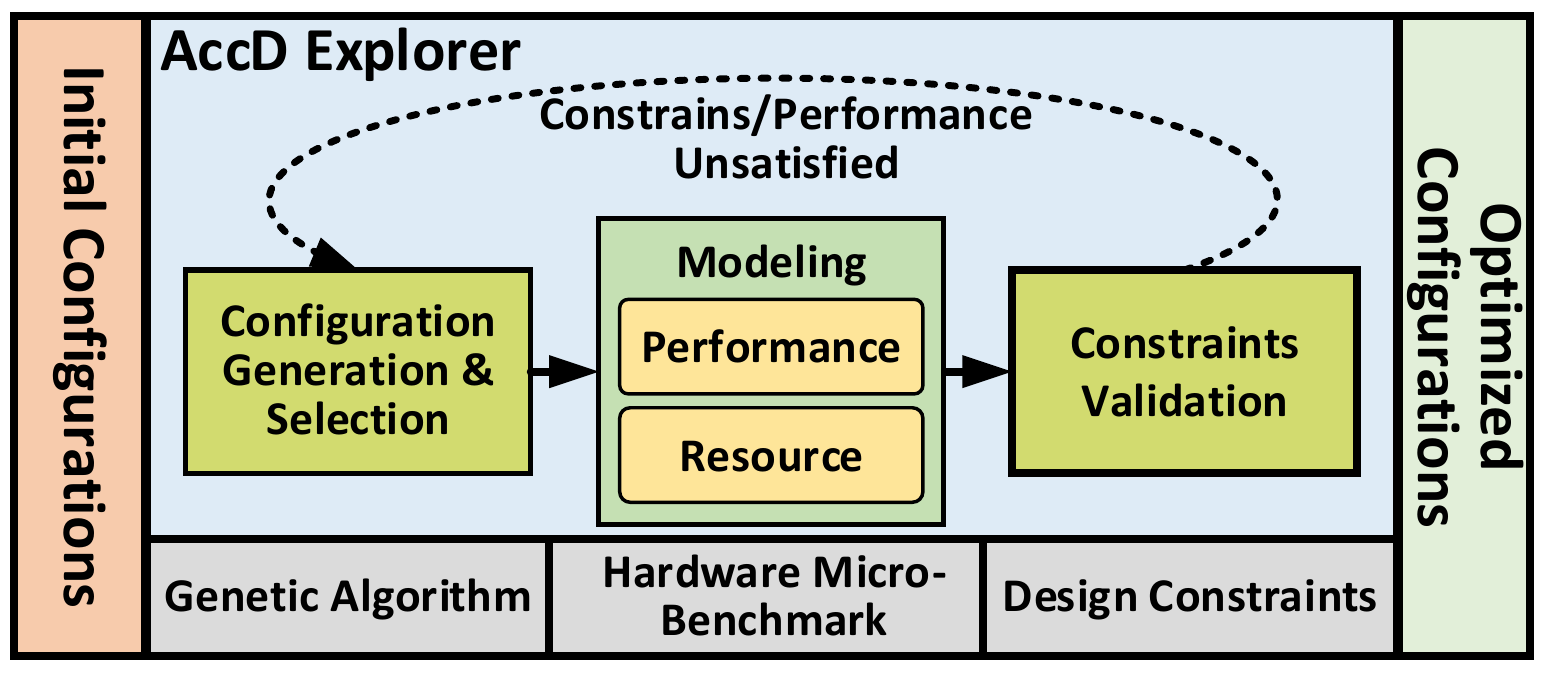}
    \caption{AccD Explorer.}
    \label{fig: AccD Explorer}
\end{figure}

Finding the best combination of design configurations (a set of hyper-parameters) under the given constraints requires non-trivial efforts in the design space search. Therefore, we incorporate an AccD explorer in our compiler framework for efficient design space exploration (Figure~\ref{fig: AccD Explorer}). AccD explorer takes a set of raw configurations (hyper-parameters) as the initial input, and generates the optimal configuration as the output through several iterations of the design configuration optimization process. In particular, AccD explorer consists of three major phases: \textit{Configuration Generation and Selection}, \textit{Performance and Resource Modeling}, \textit{Constraints Validation}. 
\vspace{-1.3em}
\paragraph{Configuration Generation and Selection}
The functionality of this phase depends on its input. There are two kinds of inputs: If the input is from the initial configurations, this phase will directly feed these configurations to the modeling phase for performance and resource evaluation; If the input is the result from the constraints validation in the last iteration, this phase will leverage the genetic algorithm to crossover the "premium" configurations kept from the last iteration, and generate a new set of configurations for the modeling phase. 
\paragraph{Performance Modeling}
Performance modeling measures the design latency and bandwidth requirement based on the input design configurations. We formulate the design latency by using Equation~\ref{equ: Latency model},
\begin{equation} \small
    \label{equ: Latency model}
     \begin{multlined}
        Latency = Latency_{filt} + Latency_{comp}
    \end{multlined}
\end{equation}
where $Latency_{filt}$ and $Latency_{comp}$ are the time of the GTI filtering process and remaining distance computations, respectively. And they can be calculated as Equation~\ref{equ: Runtime Filtering.},
\begin{equation} \small
\label{equ: Runtime Filtering.}
\begin{aligned}
    Latency_{filt} &=\frac{n_{trg\_grp}\times n_{src\_grp} \times src_{size}\times trg_{size}\times d}{n_{iteration}} 
    \\
    Latency_{comp} &=\frac{src_{size}\times trg_{size}\times ratio_{save}\times d}{blk^2\times frequency\times unroll \times simd}
\end{aligned}
\end{equation}
where $n_{src\_grp}$ and $n_{trg\_grp}$ are the number of groups for source and target points, respectively; $src_{size}$ and $trg_{size}$ are the number of points inside source and target set, respectively; $d$ is the data dimensionality; $n_{iteration}$ is the number of grouping iteration; $blk$ is the size of computation kernel block; $frequency$ is the FPGA design clock frequency; $unroll$ is the distance computation unroll factor; $simd$ is the number of parallel worker threads inside each computation block; $ratio_{save}$ is the distance saving ratio through GTI filtering (Equation~\ref{equ: Saving Ratio}),
\begin{equation} \small
    \label{equ: Saving Ratio}
    ratio_{save} = \frac{n_{iteration}}{\alpha} \times \sqrt{\frac{src_{size}\times trg_{size}}{n_{src\_grp} \times n_{trg\_grp}}}
\end{equation}
where the $\alpha$ the density of points distribution. This formula also tells that increasing of number of iterations and number of points inside each group would improve the performance of GTI filtering performance. Also, the increase of points distribution density $\alpha$, (\textit{i.e.} points are closer to each other) will decrease the GTI filtering performance.

To get the required bandwidth $BW$ of the current design, we leverage Equation~\ref{equ: required bandwith}, 
\begin{equation} \small
\label{equ: required bandwith}
    BW = \frac{(src_{size} + trg_{size})\times d \times size_{data\_type}}{Latency}
\end{equation}
where the $size_{data\_type}$ can be either 32-bit for \textit{int} and \textit{float} or 64-bit for \textit{double}. 

\paragraph{Resource Modeling}
Directly measuring the hardware resource usage of the accelerator design from high-level algorithm description is challenging because of the hidden transformation and optimization in hardware design suite. However, AccD uses a micro-benchmark based methodology to measure the hardware resource usage by analytical modeling. The major hardware resource consumption of AccD comes from the distance computation kernel, which depends on several design factors, including the kernel block size, the number of SIMD workers, etc.

In AccD resource analytical model, the design factors are classified into two categories: \textit{dataset-dependent} and \textit{dataset-independent} factors. The main idea behind the AccD resource modeling is to get the exact hardware resource consumption statistics through micro-benchmark on the hardware designs with different dataset-independent factors. For example, we can benchmark a single distance computation kernel with different sizes of computation block to get its resource statistics. Since this factor is dataset-independent, which can be decided before knowing the dataset details. However, to estimate the resource consumption for datasets with different sizes and dimensionalities, AccD leverages the formula-based approach to estimate the overall hardware resource consumption (Equation~\ref{equ: Resource estimation}), which combines online information (\textit{e.g.}, kernel organization, and dataset properties) and offline information (\textit{e.g.}, miro-benchmark statistics).
\begin{equation} \small
    \label{equ: Resource estimation}
    \begin{aligned}
         Resource_{est} = Resource_{single}\times \mathbf{ceil}(\frac{src_{size}}{blk})\times \mathbf{ceil}(\frac{trg_{size}}{blk}) \\
    \end{aligned}
\end{equation}
where the types of $Resource$ can be on-chip memory, computing units, and logic operation units; $Resource_{est}$ is the estimated overall usage of a certain type resource for the overall design; $Resource_{single}$ is the usage of a certain type of resource for only one distance computation kernel block.

\paragraph{Constraints Validation}
Constraints validation is the third phase of AccD explorer, which checks whether the design resources consumption of a given configuration is within the budget of the given hardware platform. The input of this phase are the resource estimation results from resource modeling step. The design constraint inequalities are listed in Equation~\ref{eqn: Resource Constraints}, which includes $Mem$ (the size of on-chip memory), $BW$ (the bandwidth of data communication between external memory and on-chip memory), $Computing\_Unit$ (the number of computing units) and $Logic\_Unit$ (the number of logic units):
\begin{equation} \small
\label{eqn: Resource Constraints}
\begin{aligned}
    BW &\leq BW_{max}, 
    \\
    Mem &\leq Mem_{max}, 
    \\
    Computing\_Unit &\leq  Computing\_Unit_{max},
    \\
    Logic\_Unit &\leq  Logic\_Unit_{max}
\end{aligned}
\end{equation}

Constraints validation phase will also discard the configurations that cannot match the design performance and constraints, and only keep the "well-performed" configurations for further optimization in the next iteration. The constraints validation phase will also record the modeling information of the best configuration  statistics in the last iteration, which will be used to terminate the optimization process if the modeling results difference between the configurations in two consecutive iterations is lower than a predefined threshold. This strategy can also help to avoid unnecessary time cost. After termination of the AccD explorer, the "best" configuration with maximum design performance under the given constraints will be output as the "optimal" solution for the AccD design. 

\newcolumntype{L}[1]{>{\raggedright\arraybackslash}p{#1}}
\newcolumntype{C}[1]{>{\centering\arraybackslash}p{#1}}
\newcolumntype{R}[1]{>{\raggedleft\arraybackslash}p{#1}}

\section{Evaluation} \label{sect: Evaluation}
In this section, we choose three representative benchmarks (K-means, KNN-join, and N-body Simulation) and evaluate their corresponding AccD designs on the CPU-FPGA platform.

\paragraph{K-means}
K-means~\cite{LloydKMeans,dataclustering50, efficientKmeans, coates2012learning, ray1999determination} clusters a set of points into several groups in an iterative manner. At each iteration, it first computes the distances between each point and all clusters, and then update the clusters based on the average position of their inside points. We choose it as our benchmark since it can show the  benefits of AccD hierarchy (\textbf{Trace-based + Group-level}) bound computation optimization on iterative algorithms with disjoint source and target set.

\paragraph{KNN-join}
KNN-join Search~\cite{altman1992introduction, KNNJoinsHybridApproach, KNNJoinsDataStreams} finds the Top-K nearest neighbor points for each point in the source set from the target set. It first computes the distances between each source point and all the target points. Then it ranks the K-smallest distances for each source point and gets its corresponding closest Top-K target points. KNN-join can help to demonstrate the effectiveness of AccD hybrid (\textbf{Two-landmark + Group-level}) bound computation optimization on non-iterative algorithms. 

\paragraph{N-body Simulation}
N-body Simulation~\cite{nylons2007fast, ida1992n} mimics the particle movement within a certain range of 3D space. At each time step, distances between each particle and its neighbors (within a radius $R$) are first computed, and then the acceleration and the new position of each particle will be updated based on these distances. While N-body simulation is also iterative, it has several differences compared with K-means algorithm: 1) N-body simulation has the same dataset (particles) for source and target set, whereas K-means operates on different source (point) and target (cluster) sets; 2) All points in the N-body simulation would change their positions according to the time variation, whereas in K-means only the target set (cluster) would change their positions during the center update; 3) N-body simulation has the same size of source and target set, whereas K-means target set (cluster) is much smaller than source set (point) in general. N-body simulation can help us to show the strength of AccD hybrid bound computation (\textbf{Two-landmark + Trace-based + Group-level}) on iterative algorithms with the same source and target set.

\vspace{-0.3\baselineskip}
\subsection{Experiment Setup}
\vspace{-0.2\baselineskip}
\paragraph{Tools and Metrics} 
In our evaluation, we use Intel Stratix 10 DE10-Pro~\cite{DE10-Pro} as the FPGA accelerator and run the host side software program on Intel Xeon Silver 4110 processor~\cite{intelXeon} (8-core 16-thread, 2.1GHz base clock frequency, 85W TDP). DE10-Pro FPGA has 378,000 Logic elements (LEs), 128,160 adaptive logic modules (ALM), 512,640 ALM registers, 648 DSPs, and 1,537 M20K memory blocks. We implement AccD design on DE10-Pro by using Intel Quartus Prime Software Suite~\cite{IntelQuartus} with Intel OpenCL SDK included. To measure the system power consumption (Watt) accurately, we use the off-the-shelf Poniie PN2000 as the external power meter to get the runtime power of Xeon CPU and DE10 Pro FPGA.
\vspace{-1em}
\begin{table}[h] \small
\centering
\caption{Implementation Description.}
\label{table: Implementation Description.}
\begin{tabular}{|| c | C{9em} | L{9em} ||}
\hline
\textbf{Name}
& \makecell{\ \textbf{Techniques}}
& \makecell{\ \textbf{Description}} 
\\ 
\hline
\hline  \textbf{Baseline}   & Standard Algorithm without any optimization, CPU. & Naive for-loop based implementation on CPU.
\\ 
\hline  \textbf{TOP}   & Point-based Triangle-inequality Optimized Algorithms, CPU. & TOP \cite{Topframework} optimized distance-related algorithm running on CPU.
\\ 
\hline  \textbf{CBLAS} & CBLAS library Accelerated Algorithms, CPU. & Standard distance-related algorithm with CBLAS~\cite{openblas} acceleration.
\\
\hline  \textbf{AccD} & Algorithmic-hardware co-design, CPU-FPGA platform. & GTI filtering and FPGA acceleration of distance computations.
\\
\hline
\end{tabular}
\end{table}
\begin{table*}[t] \small
\centering
 \begin{tabular}{|| l c c c || l c  c || c c ||} 
\hline
 \multicolumn{4}{||c||}{\textbf{K-means}} 
 & \multicolumn{3}{c||}{\textbf{KNN-join}} 
 & \multicolumn{2}{c||}{\textbf{N-body Simulation}}\\  
\hline
 \textbf{Dataset} & \textbf{Size} & \textbf{Dimension} & \textbf{\#Cluster} 
 &
 \textbf{Dataset} & \textbf{Dimension}  & \textbf{\#Source} & 
 \textbf{Dataset} & \textbf{\#Particle} \\
\hline
Poker Hand &   25,010   &	11	    &    158   &  
Harddrive1	&  64       &   68,411  &
P-1 & 16,384  
 \\ 
Smartwatch Sens        &   58,371	& 12	&   242     &  
Kegg Net Directed	   &  24        & 53,413	 &
P-2 & 32,768
\\
Healthy Older People    &   75,128 & 	9	    & 274       &  
3D Spatial Network	    &   3     &    434,874    &
P-3 & 59,049
\\
KDD Cup 2004  &  285,409    & 	74	    &   534     &  
KDD Cup 1998	    & 56        & 95,413	 &
P-4 & 78,125    
\\
Kegg Net Undirected  &  65,554 &	28	& 256   & 
Skin NonSkin        &   4 & 245,057  & 
P-5 & 177,147  
\\
Ipums       &  70,187	& 60    &	265     &  
Protein	    &  11       & 26,611 &
P-6 & 262,144  
 \\
 \hline
 \end{tabular}
     \caption{Datasets for Evaluation.}
     \label{table: Evaluation Dataset}
\end{table*} 

\begin{figure*}[h] \small
    \centering
    \subfloat[]{\includegraphics[width=0.33\textwidth]{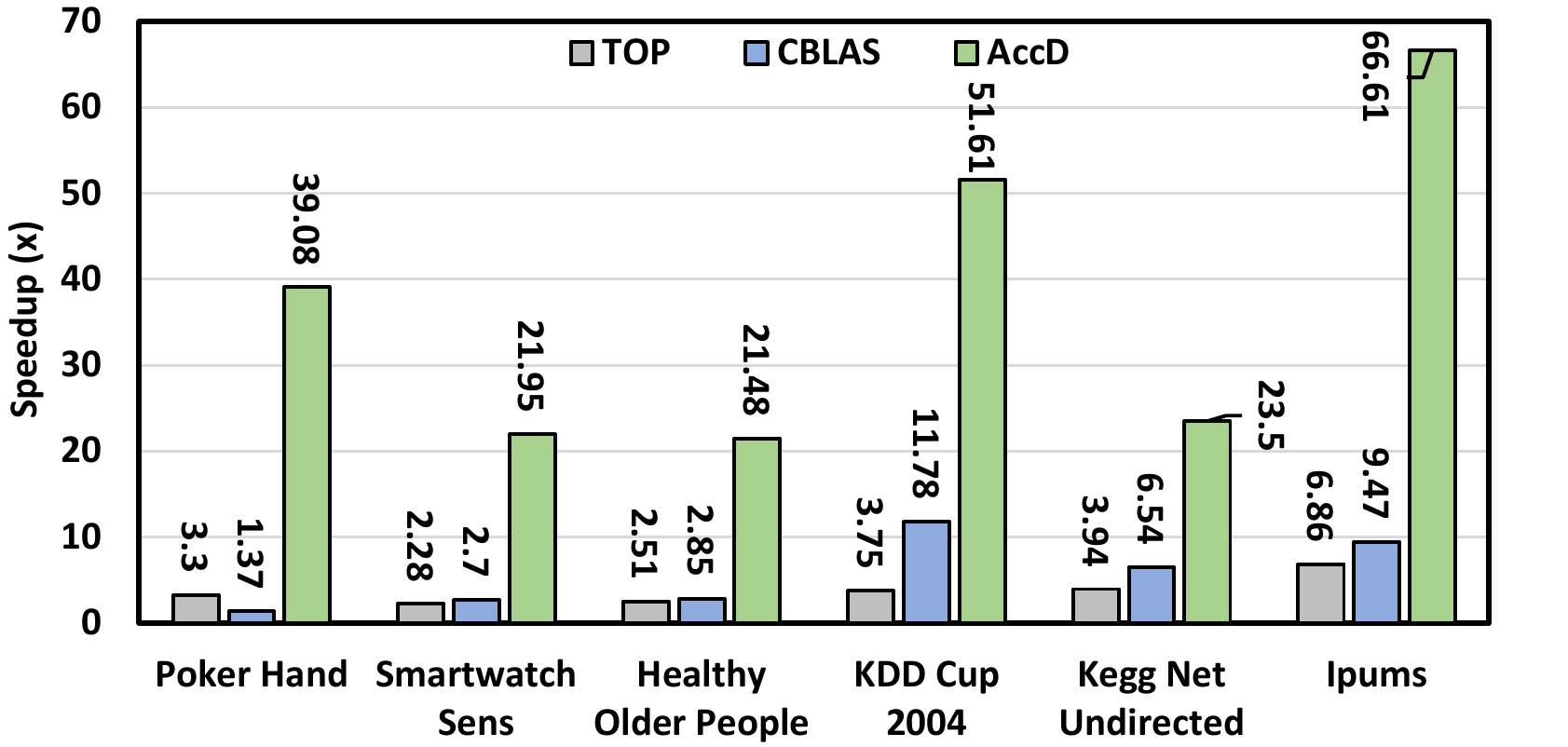}}
    \subfloat[]{\includegraphics[width=0.33\textwidth]{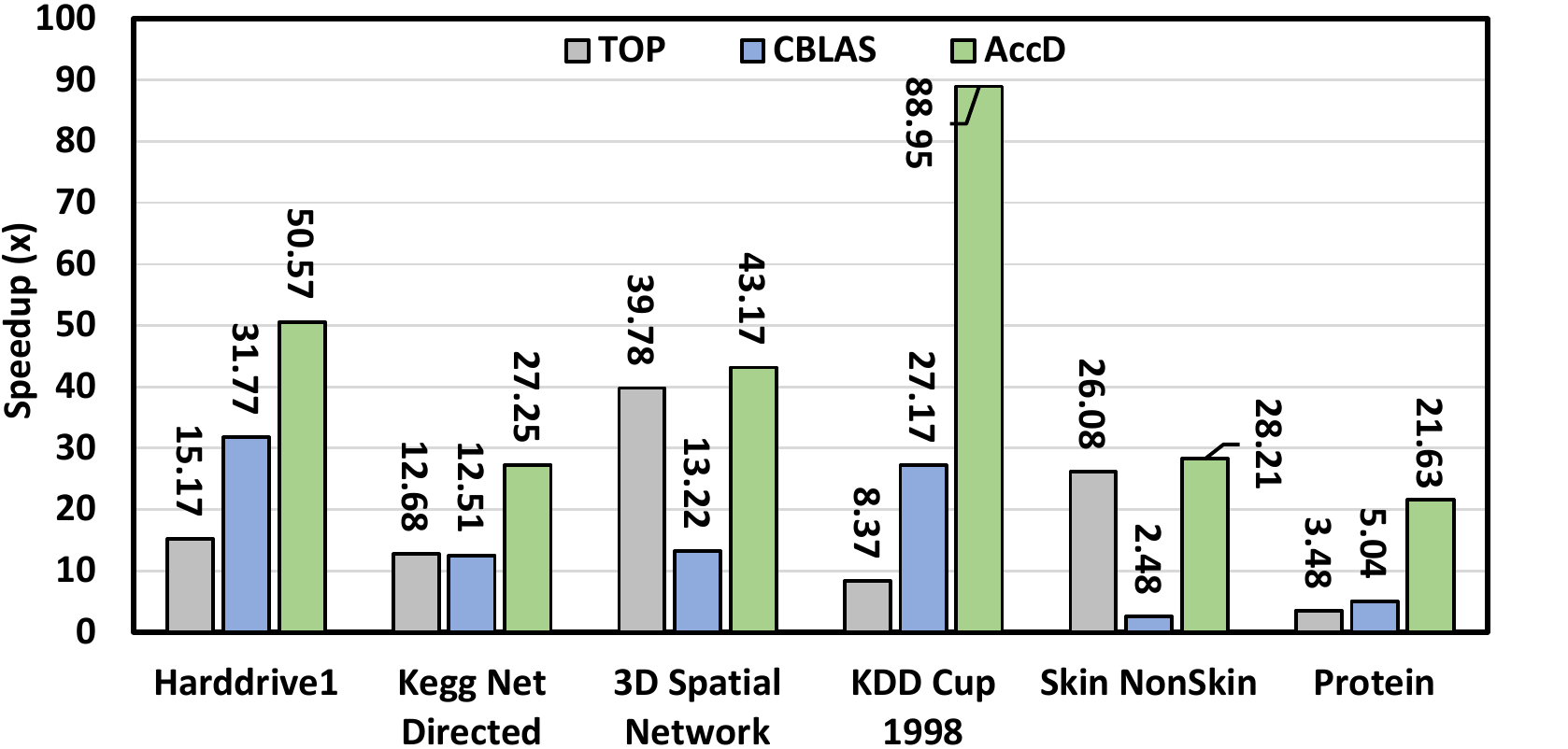}}
    \subfloat[]{\includegraphics[width=0.33\textwidth]{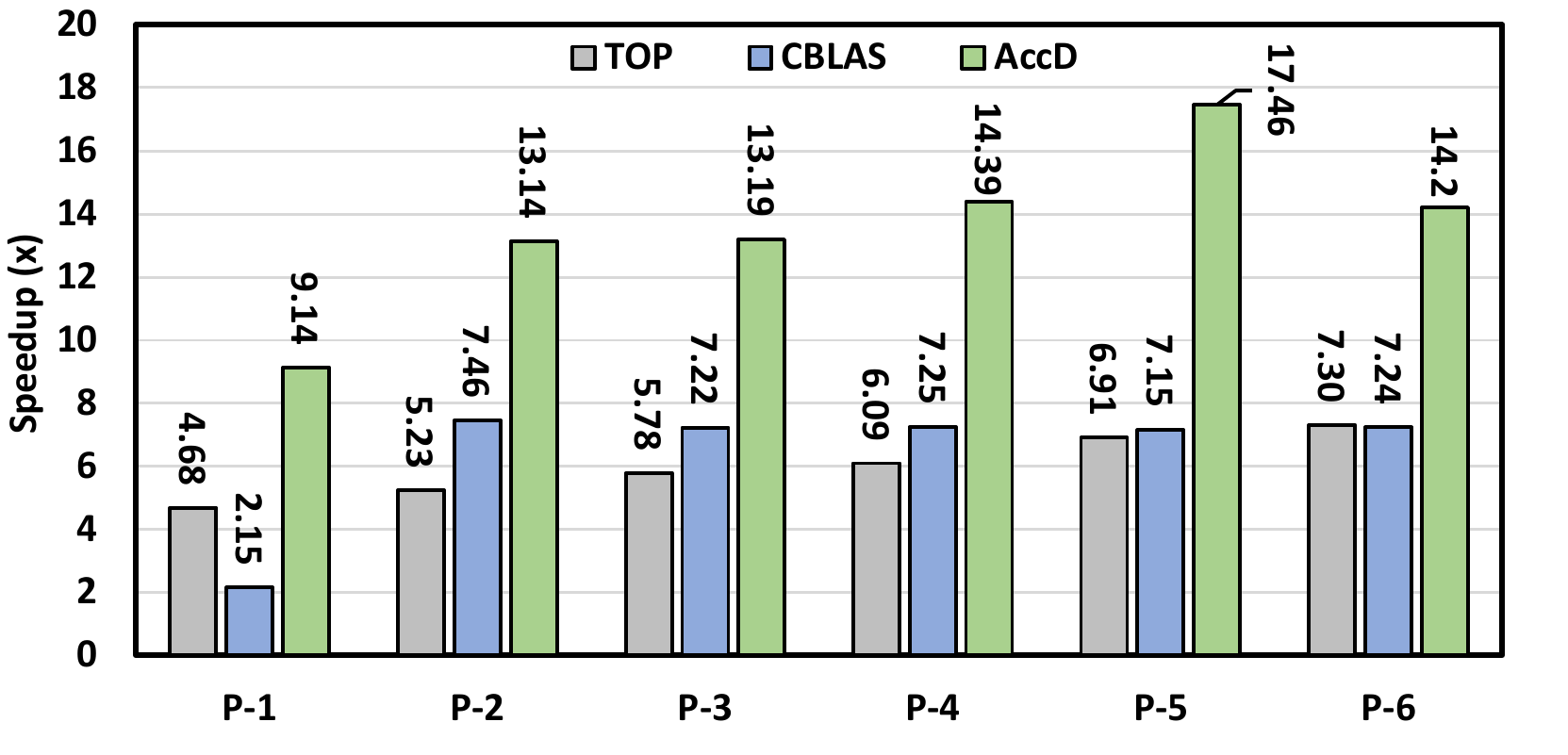}}
    \caption{Performance Comparison (TOP, CBLAS, AccD): (a) K-means (b) KNN-Join (c) N-body Simulation. Note: Speedup is normalized w.r.t Baseline.}
    \label{fig: Speedup Performance Comparison.}
\end{figure*} 
\begin{figure*}[ht!] \small
    \centering
    \subfloat[]{\includegraphics[width=0.33\textwidth]{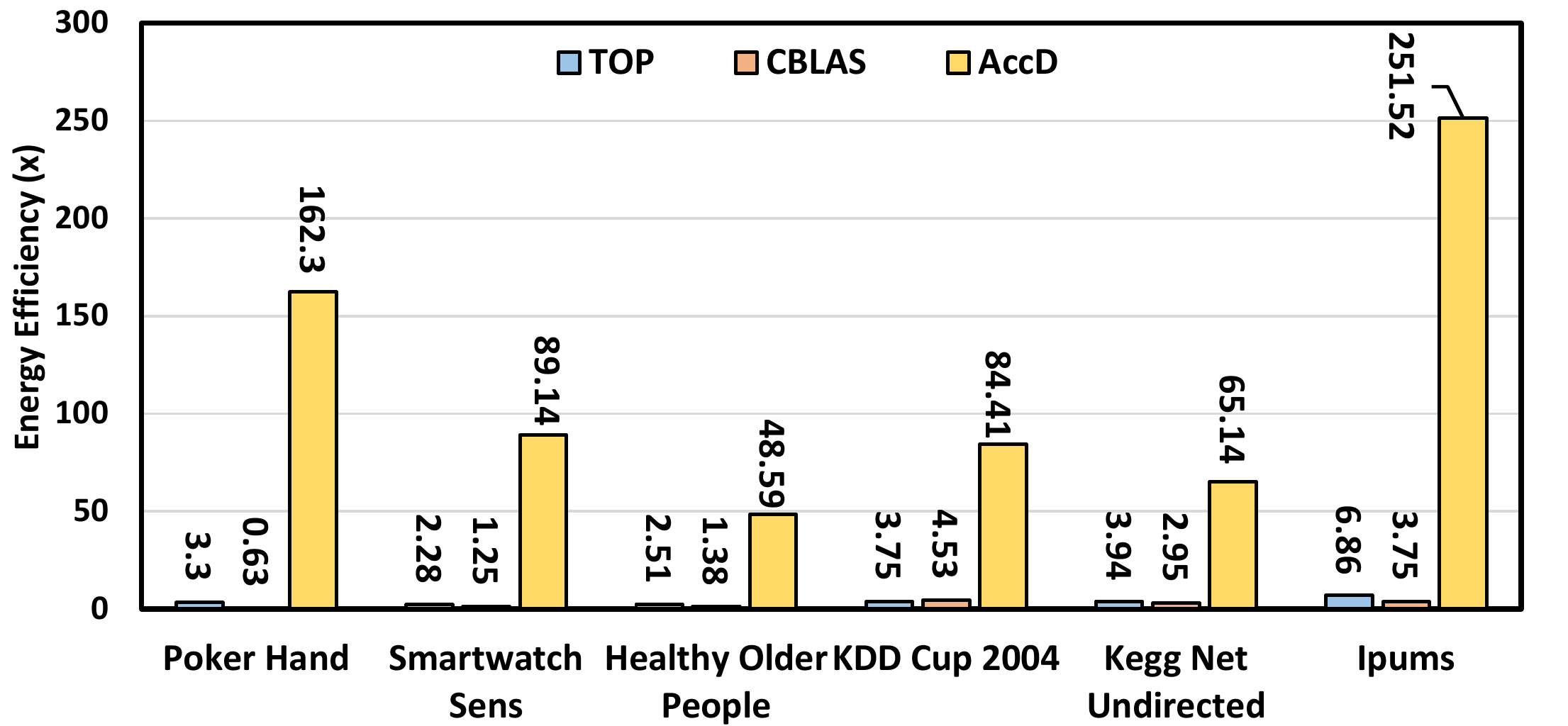}} 
    \subfloat[]{\includegraphics[width=0.33\textwidth]{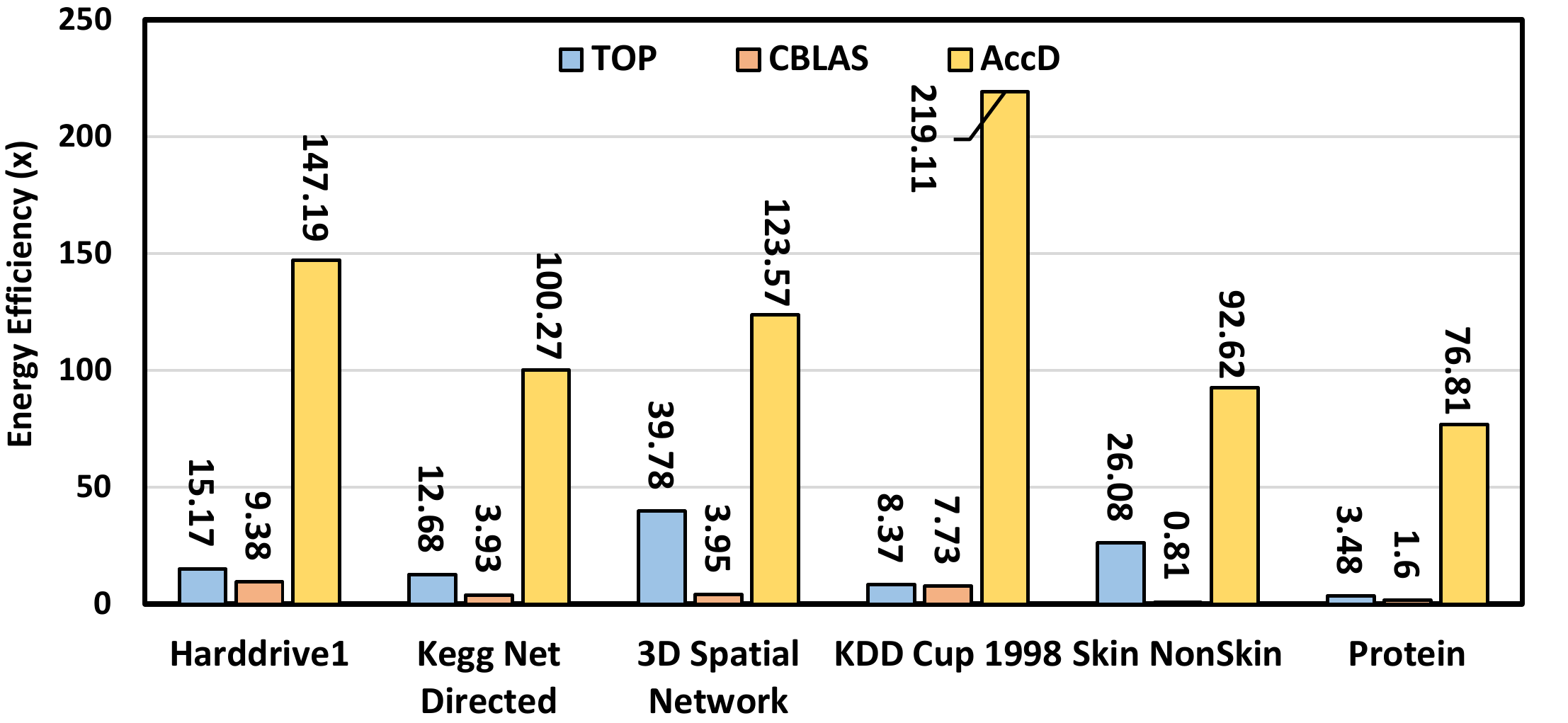}} 
    \subfloat[]{\includegraphics[width=0.33\textwidth]{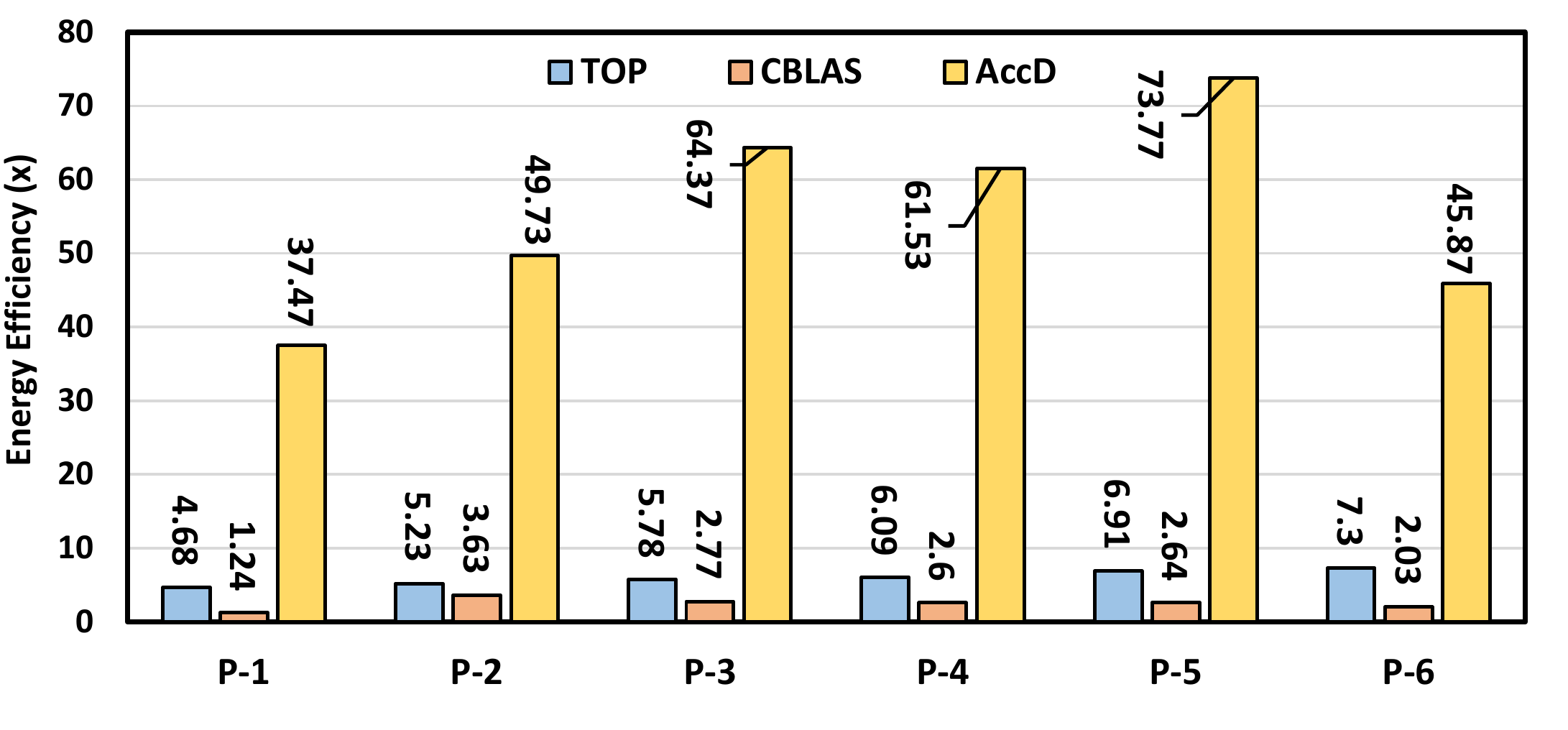}}
    \caption{Energy Efficiency Comparison (TOP, CBLAS, AccD): (a) K-means. (b) KNN-Join. (c) N-body Simulation. Note: Energy Efficiency is normalized w.r.t Baseline.}
    \label{fig: Energy Efficiency Comparison}
\end{figure*}
\vspace{-1em}
\paragraph{Implementations} 
The CPU-based implementations consist of three types of programs: the naive for-loop sequential implementation without any optimization (selected as our \textbf{Baseline} to normalize the speedup and energy-efficiency), the algorithm optimized by \textbf{TOP}~\cite{Topframework} framework and the algorithm optimized by \textbf{CBLAS}~\cite{openblas} computing library. Note that the TOP + CBLAS implementation is \textbf{not} included in our evaluation, since after applying TOP point-based TI filtering, each point in the source set has a distinctive list of points from the target set for distance computation, whereas CBLAS requires uniformity in the distance computations. Therefore, it is challenging to combine TOP and CBLAS optimization.

\paragraph{Dataset} 
In the evaluation, we use six datasets for each algorithm. The selected datasets can cover the wide spectrum of mainstream datasets, including datasets from UCI Machine Learning Repository~\cite{UCI-dataset}, and datasets that have ever been used by previous papers~\cite{Topframework, ding2015yinyang, chen2017sweet} in the related domains. Details of these datasets are listed in Table~\ref{table: Evaluation Dataset}. Note that KNN-join algorithm will find the Top-1000 closest neighbors of each query point.

\subsection{Comparison with Software Implementation} 
\vspace{-0.2\baselineskip}
\label{sect: Comparison with Software Implementation}
\paragraph{Performance Comparison} 
As shown in Figure~\ref{fig: Speedup Performance Comparison.}, TOP, CBLAS, and AccD achieve average $9.12\times$,  $9.19\times$ and $31.42\times$ compared with Baseline across all algorithm and dataset settings, respectively. As we can see, AccD design can always maintain the highest speedup among these implementations. This largely dues to AccD GTI optimization in reducing distance computation and its efficient hardware acceleration of the distance computation on FPGA.

We also observe that TOP implementation shows its strength for large datasets. For example, on dataset 3D Spatial Network ($n=434,874$) in KNN-join, TOP implementation achieves $39.78\times$ speedup. Since the fine-grained point-based TI optimization of TOP can reduce most (more than 90\%) of the unnecessary distance computations, which benefits the overall performance to a great extent. Note that the intrinsic point distribution of the dataset would also affect the filtering performance of TOP, but in general, the larger dataset could lead TOP to spot and remove more redundant computations.

What we also notice is that CBLAS implementation demonstrates its performance on datasets with relatively high dimensionality. For example, on dataset KDD Cup 2004 ($d=74$) in the K-means algorithm, CBLAS achieves $11.78\times$ speedup over Baseline, which is higher than its performance on other K-means datasets. This is because, on high dimension dataset, CBLAS implementation can get more benefits of parallelized computing and more regularized memory access, whereas, in low dimension settings, the same optimization can only yield minor speedup.

Our AccD design achieves a considerable speedup on datasets with large size and high dimensionality. For example, on dataset KDD Cup 2004 ($n=285,409, d=74$) and Ipums ($n=70,187, d=60$) in K-means, AccD achieves $51.61\times$ and $66.61\times$ speedup over Baseline, and also significantly higher than both TOP and CBLAS implementations. This conclusion can also be extended to KNN-join, such as $88.95\times$ speedup on dataset KDD Cup 1998 ($n=95,413$, $d=56$). Since our AccD design can effectively reconcile the benefits from both the GTI optimization and the FPGA acceleration, where the former provides the opportunity to reduce the distance computation at the algorithm level, and the latter boosts the performance from hardware acceleration perspective. More importantly, our AccD design can balance the above two benefits to maximize the overall performance. 
\paragraph{Energy Comparison} 
The energy efficiency of AccD design is also significant. For example, on the K-means algorithm, AccD designs deliver an average $116.85\times$ better energy efficiency compared with Baseline, which is significantly higher than TOP and CBLAS implementations. There are namely two reasons behind these results: 1) Much lower power consumption. AccD CPU-FPGA design only consumes $5w \sim 17.12w$ across all algorithm and dataset settings, whereas Intel Xeon CPU consumes at least $20.9w$ and $42.49w$ on TOP and CBLAS implementations, respectively; 2) Considerable performance. AccD design achieves a much better speedup (more than $5\times$ on average) compared with the TOP and CBLAS, which contributes to overall design energy-efficiency. 

Among these implementations, CLBAS implementation has the lowest energy efficiency, since it relies on multi-core parallel processing capability of the CPU, which improves the performance at the cost of much higher power consumption (average $65.79w$). TOP only leverages the single-core processing capability of the CPU and achieves moderate performance with effective distance computation reduction, which results in less power consumption (average $25.59w$) and higher energy efficiency (average $9.12\times$) compared with Baseline. Different from the TOP and CBLAS implementations, AccD design is built upon a low-power platform with considerable performance, which shows a far better energy-performance trade-off. 

\vspace{-0.15\baselineskip}
\subsection{Performance Benefits Analysis}
\vspace{-0.15\baselineskip}
To analyze the performance benefits of AccD CPU-FPGA design in detail, we use K-means as the example algorithm for study. Specifically, we build four implementations for comparison: 1) TOP K-means on CPU; 2) TOP K-means on CPU-FPGA platform; 3) AccD K-means on CPU; 4) AccD K-means on CPU-FPGA platform. Note that TOP K-means is designed for sequential-based CPUs, and no publicly available TOP implementation on CPU-FPGA platforms. For a fair comparison, we implement TOP K-means on CPU-FPGA platform with memory optimizations (inter-group and intra-group memory optimization) and distance computation kernel optimization (Vector-Matrix multiplication).  These optimizations improve the data reuse and memory access performance.
\vspace{-0.8\baselineskip}
\begin{figure} [ht] \small
    \centering
    \makebox{\includegraphics[width=0.85\columnwidth]{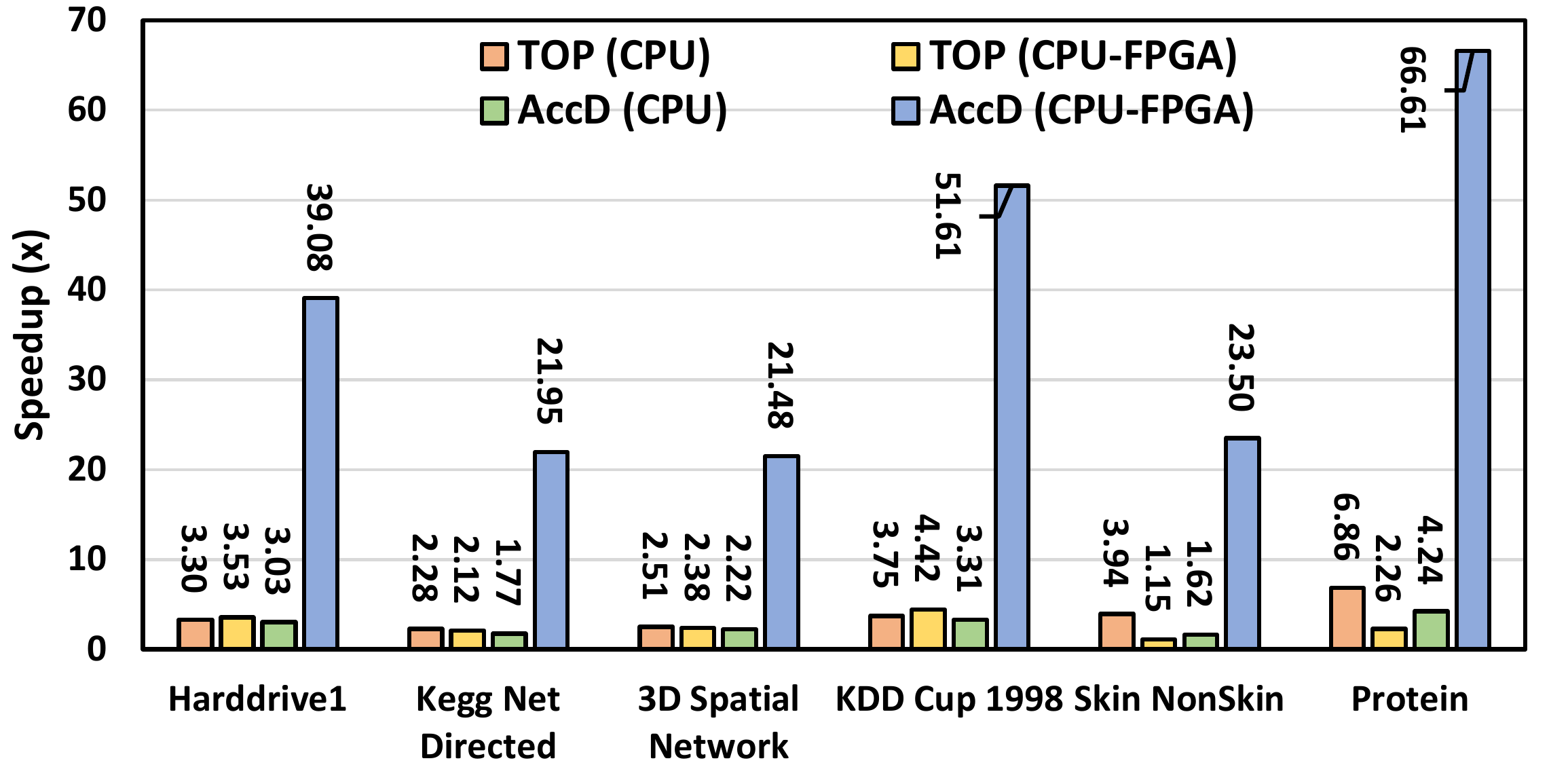}}
    \caption{AccD Performance Benefits Breakdown.}
    \label{fig: AccD Performance Benefits Breakdown.}
\end{figure}
We compute the normalized speedup performance of each implementation w.r.t the naive for-loop based K-means implementation on CPU. 

As shown in Figure~\ref{fig: AccD Performance Benefits Breakdown.}, AccD K-means on CPU-FPGA platform can always deliver the best overall speedup performance among these implementations. We also observe that TOP K-means can achieve average $3.77\times$ speedup on CPU, however, directly porting this optimization towards CPU-FPGA platform could even lead to inferior performance (average $2.63\times$). Even though we manage to add several possible optimizations, applying such fine-grained TI optimization from TOP would still cause a large divergence of computation among points, leading to low data reuse and inefficient memory access. 

We also notice that AccD design on CPU achieves lower speedup (average $2.69\times$) compared with the TOP (average $3.77\times$), since its coarse-grained GTI optimization spots a fewer number of unnecessary distance computations. However, when combining AccD design with CPU-FPGA platform, the benefits of AccD GTI optimization become prominent (average $37.37\times$), since it can maintain computation regularity while reducing memory overhead to facilitate the hardware acceleration on FPGA. Whereas, applying optimization to maximize the algorithm-level benefits while ignoring hardware-level properties would result in poor performance, such as the TOP (CPU-FPGA) implementation. 
Moreover, comparison of AccD (CPU) and AccD (CPU-FPGA) can also demonstrate the effectiveness of using FPGA as the hardware accelerator to boost the performance of the algorithms, which can deliver additional $9.68\times \sim 15.71\times$ speedup compared with the software-only solution.
\vspace{-0.3\baselineskip}
\section{Conclusion} 
\vspace{-0.4\baselineskip}
\label{sect: Conclusion}
In this paper, we present our AccD compiler framework to accelerate the distance-related algorithms on the CPU-FPGA platform. Specifically, AccD leverages a simple but expressive language construct (DDSL) to unify the distance-related algorithms, and an optimizing compiler to improve the design performance from algorithmic and hardware perspective systematically and automatically.
Rigorous experiments on three popular algorithms (K-means, KNN-join, and N-body simulation) demonstrate the AccD as a powerful and comprehensive framework for hardware acceleration of distance-related algorithms on the modern CPU-FPGA platforms.

\bibliographystyle{unsrtnat}
\bibliography{references}

\end{document}